\documentclass[useAMS,usenatbib]{mn2e}
\usepackage{graphicx}
\usepackage{latexsym}
\usepackage{url}
\usepackage{longtable}

\title{The 
  $\dot{M}-M_*$ relation of pre-main sequence stars: a consequence of
  X-ray driven disc evolution}

\author[Barbara Ercolano, Daniel Mayr, James Owen, Giovanni Rosotti, 
Carlo Felice Manara]{B. Ercolano$^{1,2}$\thanks{E-mail:
    ercolano@usm.lmu.de (BE)}, D. Mayr$^{1}$, J. E. Owen$^{3}$, G.
  Rosotti$^{1,2}$, C. F. Manara$^{4}$\\
$^{1}$Universit\"ats-Sternwarte M\"unchen, Scheinerstr. 1, 81679 M\"unchen, Germany\\
$^{2}$Excellence Cluster Origin and Structure of the Universe,
Boltzmannstr.2, 85748 Garching bei M\"unchen, Germany\\
$^{3}$Canadian Institute for Theoretical Astrophysics, 60 St. George Street, Toronto, M5S 3H8, Canada\\
$^{4}$European Southern Observatory, Karl-Schwarzschild-Str. 2, 85748 Garching bei M\"unchen, Germany}

\begin{document}

\pagerange{\pageref{firstpage}--\pageref{lastpage}} \pubyear{2011}

\maketitle

\label{firstpage}

\def\mnras{MNRAS}
\def\apj{ApJ}
\def\aap{A\&A}
\def\apjl{ApJL}
\def\apjs{ApJS}
\def\bain{BAIN}
\def\araa{ARA\&A}
\def\pasp{PASP}
\def\aj{AJ}
\def\pasj{PASJ}
\def\ga{\sim}
\voffset-.4in

\begin{abstract}

We analyse current measurements of accretion rates onto pre-main sequence
stars as a function of stellar mass, and conclude that the steep
dependance of accretion rates on stellar mass is real and not driven
by selection/detection threshold, as has been previously feared. 

These conclusions are reached by means of statistical tests including a
survival analysis which can account for upper limits. 
The power-law slope of the $\dot{M}-M_*$ relation is found to be in
the range of 1.6-1.9 for young stars with masses lower than
1~$M_{\odot}$. 

The measured slopes and distributions can be easily reproduced by
means of a simple disc model which includes viscous accretion and
X-ray photoevaporation. 
We conclude that the $\dot{M}-M_*$ relation in pre-main sequence stars
bears the signature of disc dispersal by X-ray photoevaporation,
suggesting that the relation is a straightforward consequence of disc
physics rather than an imprint of initial conditions.

\end{abstract}


\section{Introduction}
The scaling of the accretion rate, $\dot{M}$, with stellar mass, $M_*$ for
low-mass stars has been the focus of much debate over the last few
years. Measurements in the first half of the naughties, indicated that
$\dot{M}$ correlates with the square of the stellar mass (Muzerolle et
al 2003; Natta et al 2004; Calvet et al 2004; Muzerolle et al 2005;
Mohanty et al 2005; Natta et al 2006). This deviation 
from a simple linear scaling encouraged the development of a number of
theoretical models to interpret this results, including Bondi-Hoyle
accretion (Padoan et al 2005, see also Throop \& Bally 2008) and
dependance on the initial conditions 
of the parent cloud from which the protoplanetary disc formed
(Dullemond, Natta \& Testi, 2006, Alexander \& Armitage 2006). Clarke
\& Pringle (2006, CP06), and 
later Tilling et al. (2008), questioned
the quantitative value of the power-law exponent, $\alpha$, in the $\dot{M}$-$M_*$ relation,
suggesting that incompleteness of the data at both high and low
accretion rates may have conspired to yield a higher than expected
value of $\alpha$. By considering disc dispersal by EUV
photoevaporation CP06 derive a theoretical value of $\alpha = 1.35$. A
similar slope was also obtained by an independent model of Gregory et al (2006)
based on a steady state accretion which considered both dipolar and
complex magnetic fields. 

Recent observational data, however, lend credence to higher values of
the $\alpha$ exponent. 
Using different observational methods and samples in different
  regions, the typical derived values of $\alpha$ are around 1.5-1.8
  (e.g.,  Herczeg \& Hillenbrand 2008; Rigliaco et al. 2011a;
  Antoniucci et al. 2011; Biazzo et al. 2012). In particular, Manara
et al. (2012) used the Hubble Space 
Telescope to investigate the $\dot{M}$-$M_*$ relation in the Orion
Nebula Cluster, finding a value of $\alpha = 1.68 \pm$ 0.02
(compatible with the results of Natta et al. 2006). Selecting sources
according to the method used for the determination of the accretion
rates in the same sample returns values varying from  $\alpha$ = 1.59
$\pm$ 0.04 to  $\alpha$ = 1.73 $\pm$ 0.02. 

In this paper we show that values of $\alpha$~$\sim$~1.45-1.70 are
expected for stars with solar mass or lower, in the context of a
protoplanetary disc dispersal mechanism based on 
X-ray photoevaporation. The work is organised as follows. In section 2
we describe the available observational data and perform some simple
survival statistics to account for upper limit measurements. In
Section 3 we describe the theoretical prediction of the $\dot{M}$-$M_*$
relation  for a population of discs dispersed by X-ray
photoevaporation, showing that this agrees with the observational
values and perform additional statistical tests. In Section 4 we briefly summarise our findings. 

\section{Observational Samples}

We have collected mass accretion rates versus stellar mass data from
the literature in order to investigate the correlation between $\dot{M}$
and $M_*$ implied by recent observational data.  Our dataset is
described in Table~2. In order to address the concern raised by CP06, that the steepness of
the $\dot{M}$-M$_*$ relation may be driven by selection effects in the
data, we have collected, where available, all upper limits and included
them in the calculation of $\alpha$ by means of survival statistics
techniques. The total number of data-points we have collected is
3764 of which 15.4\% are upper limits. Necessarily a number of
measurements are duplicated in various sources, namely 84.1\% are
duplicates. In such cases we have taken the geometrical average of the
measurements. In cases where both measurements and upper limits are
available for an individual source we neglect the upper limits and
only use the measurements.  Binaries are a further source of
contamination, around 7\% of the objects in the total sample are in
known systems, but it is unclear how many
unknown binaries may still be left. In total we are left with 1623 measurements for individual objects of which 294 are upper limits. A plot of the full dataset is
shown in Figure~1.

 The largest from the recent
surveys is the HST/WFPC2 of Manara et al (2012, M12), which includes
measurements of $\dot{M}$ based on U-band excess and H-$\alpha$ luminosity
for approximately 700 sources in the ONC. The large and homogeneously
determined set of $\dot{M}$ obtained by M12 allowed them to draw some
important conclusions on the behaviour of $\dot{M}$ as a function of
stellar mass and stellar age. Based on the whole survey they found
$\alpha$ = 1.68, or $\alpha$ = 1.73/1.59 if only the sources with $\dot{M}$
measured from U-band excess/H$\alpha$ method are selected. 
Values of $\alpha \sim 1.6-1.8$ are often found when analyzing different
regions and using various methodologies. This is inconsistent with the
results by Fang et al. (2009) , who derive an $\alpha$=3 in their sample
of sub-solar mass targets in the Lynds 1630N and 1641 clouds in
Orion. This inconsistency is perhaps related to the different methodologies
used, in particular in the different relations between the accretion
luminosity and the line luminosity and
in the different evolutionary models used with respect to any other
work. When converting the values of $L_{\rm acc}$ reported by
Fang et al. (2009, 2013) in $\dot{M}$ using classical evolutionary models we
derive values of the slope $\sim$ 2, compatible with the values reported
in other works, even if still slightly higher.


\begin{table*}
\caption{Slopes of the $\dot{M}$ versus M$_*$ relation ($\alpha$)
and of the L$_X$  versus M$_*$ relation ($\beta$).}
\centering
\begin{tabular}{ l r r r r|  r r r r r}
& \multicolumn{4}{c|}{$\alpha$} & \multicolumn{4}{c}{$\beta$}\\
 & \multicolumn{1}{c}{total sample} & \multicolumn{1}{c}{$\#$} & \multicolumn{1}{c}{Manara} & \multicolumn{1}{c|}{$\#$} &  \multicolumn{1}{c}{COUP} & \multicolumn{1}{c}{$\#$} & \multicolumn{1}{c}{Güdel} & \multicolumn{1}{c}{$\#$}\\
\hline \hline
all data				& $1.66\pm0.07$ & 1320& $1.65\pm0.14$ & 698	&$1.28\pm0.07$	&544& $ 1.38\pm0.13$ & 116 \\
'' EM method				& $1.93\pm0.07$ & 1608& $2.06\pm0.14$ & 783	&	&	&				\\
'' BJ method				& $1.61\pm0.07$ & 1608& $1.38\pm0.14$ & 783	&	&	&				\\
0.032 $M_{\odot}$ $<$ M $<$ 10 $M_{\odot}$	& $1.62\pm0.07$ & 1311& $1.65\pm0.14$ & 698	&$1.44\pm0.08$	&537	&  $ 1.38\pm0.13$ & 116 \\
'' EM method				& $1.89\pm0.08$ & 1592& $2.06\pm0.14$ & 783	&	&	&				\\
'' BJ method				& $1.61\pm0.07$ & 1592& $1.38\pm0.14$ & 783	&	&	&				\\
1.0 $M_{\odot}$ $<$ M $<$ 10 $M_{\odot}$	& $3.00\pm0.43$ & 111 & $2.24\pm1.42$ & 24	&$1.46\pm0.42$  &83	& $1.04\pm1.20$ & 18 \\
'' EM method				& $3.25\pm0.48$ & 127 & \multicolumn{2}{c|}{no upper limits}&	&				\\
'' BJ method				& $3.20\pm0.42$ & 127 & \multicolumn{2}{c|}{no upper limits}&	&				\\
0.032 $M_{\odot}$ $<$ M $<$ 1.0 $M_{\odot}$	& $1.57\pm0.10$ & 1200& $1.63\pm0.18$ & 674	&$1.71\pm0.12$  &454	&  $1.44\pm0.17$ & 98  \\
'' EM method				& $1.99\pm0.10$ & 1465& $2.19\pm0.18$ & 759	&	&	&		\\
'' BJ method				& $1.61\pm0.10$ & 1465& $1.29\pm0.18$ & 759	&	&	&		\\
\label{t:references}

\end{tabular}                                                      
\newline
\end{table*}

\begin{figure}
\begin{center}
\includegraphics[width=0.3\textwidth, angle = 90]{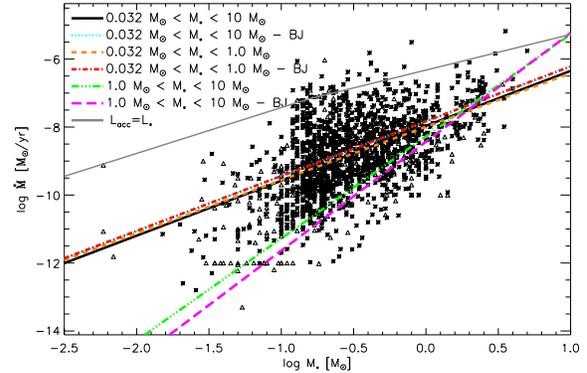}
\caption{ Accretion rates as a function of stellar mass for the whole
  collected sample. Crosses represent measurements and triangles upper
  limits. The slopes of the distributions are listed in
  Table 1. The grey solid lines shows the locus where the accretion
  and bolometric luminosity are equal.}
\label{}
\end{center}
\end{figure}


\begin{figure}
\begin{center}
\includegraphics[width=0.3\textwidth, angle = 90]{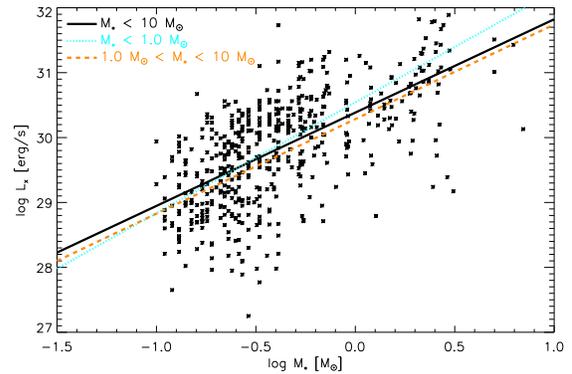}
\caption{ X-ray luminosities as a function of stellar mass in the Orion
Nebular Cluster obtained with the Chandra X-ray Telescope as part of
the COUP project (Preibisch et al 2005).
}
\label{}
\end{center}
\end{figure}

\begin{figure}
\begin{center}
\includegraphics[width=0.3\textwidth, , angle = 90]{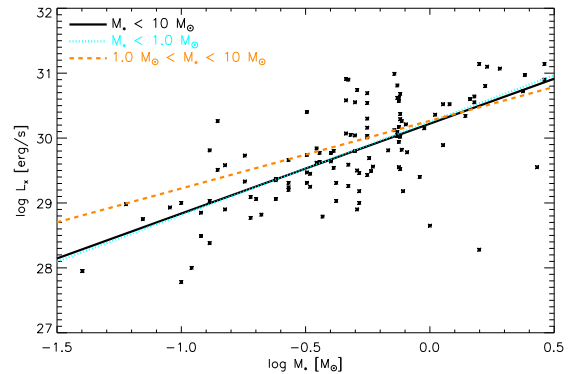}
\caption{ X-ray luminosities as a function of stellar mass in the
  Taurus Molecular Cloud obtained with the XMM Newton X-ray Telescope
  (G\"udel et al 2007).}
\label{}
\end{center}
\end{figure}

One of the main drawbacks of previous analyses and one of the
  main arguments of CP06 and Tilling et al. (2008) against over
  interpreting any possible correlation is the fact that the lowest
  possible accretion rate measurement in nearby star forming regions
  correlates strongly with stellar mass, something that is
  particularly evident in Figure~1.  Manara et al. (2013a) analyzed a sample of 24 non-accreting (Class~III)
YSOs to derive the threshold on the estimates of accretion
luminosities, L$_{acc}$,  in accreting YSOs
determined with line luminosity. They found that this threshold, when
converted in $\dot{M}$, depends on the mass of the targets. More interestingly,
this threshold happens to be right below the typical values of $\dot{M}$
derived in other works, and seems to follow the trend of the $\dot{M}$-$M_*$
relation. Given that this threshold is determined only for estimates of
L$_{acc}$ derived from line luminosity the data obtained with U-band excess
determination should not be affected. Still, the detection of U-band
excess for hotter targets is more challenging because the typical
photospheric temperature is similar to that of the accretion shock (e.g.,
Calvet et al. 2004, M12). Nevertheless, upper limit determinations should
take this effect into account.

In order to test the
  null-hypothesis that the observed correlation is purely driven by
  upper-limit selection effects we perform a Cox regression test
  (e.g. Feigelson et al. 1985) which account for the upper
  limits. Given the now large number of available measurements, we are
  able to reject the null-hypothesis at the $>$10$\sigma$ level. This
  confirms that the observed correlation between $\dot{M}$ and M$_*$ is
  infact real, and not driven by upper limit selection affects. This
  is obviously not surprising when one inspects Figure~1 given the
  amount of data now available; however, it casts away any doubt from
  previous analyses when the data samples were smaller and such a
  concern was legitimate as pointed out by CP06.

We have recalculated slopes for the $\dot{M}$-M$_*$ relation using the full
collected sample and the sample of Manara et al (2012), and summarise
the results in Table~1. We also include the
slopes obtained using two different survival statistics algorithms for
linear regression: the one assuming normal residuals (EM algorithm)
and the other assuming Kepler-Meyer residuals (The Buckley-James
algorithm, BJ). More information about the statistical methods and the
{\sc asurv} package which was employed for the analysis is given in
Lavalley, Isobe \& Feigelson (1992). The slope, $\alpha$,  of the $\dot{M}$
-M$_*$ changes significantly for different mass ranges, as already
noted by several authors (e.g. Rigliaco et al 2012). Indeed $\alpha$
is smaller in the lower stellar mass range ($M_* < 1 M_{\odot}$, $\alpha =
1.5-1.9$) compared to the higher stellar mass range ($1 M_{\odot} < M_* < 10 M_{\odot}$, $\alpha =
3.-3.2$). The dramatic change in the slope suggests that different
physical processes may be at play in the two mass
regimes. Qualitatively similar conclusions are reached by examination
of the slopes yielded by the sample of M12 alone. Typical slopes
obtained for the two mass ranges and for the full sample are shown in
Figure~1 and summarised in Table~1. 

The data used to calculate the slopes cited in Table~1 are the mass
accretion rates and stellar masses given by the various authors. This
constitutes an inhomogeneous sample as different authors
adopted different evolutionary models. We have also recalculated the
entire dataset using consistent evolutionary tracks for all objects
when possible, and note here that the slopes are sensitive to the
choice of evolutionary track. In all cases however the slopes for the
low-mass range are between 1.3 and 1.9 when upper limits are excluded
and between 1.4 and 2.4 using the Buckley-James algorithm.  The tracks
used include D'Antona \& Mazzitelli (1997), Baraffe et al. 1998, Palla
\& Stahler (1999), Siess et al. (2000, with and without overshoot).

 Another possible selection effect discussed by CP06 is that the upper
bound of the distribution is limited by those cases where the accretion
luminosity, L$_{acc}$, becomes larger than the bolometric luminosity
of the star, L$_{bol}$. Indeed this is a legitimate concern when one
analyses the data that were available in 2006. In Figure 4 of CP06 one
sees indeed that the data seem to uniformly fill the space up to k =
1, where k = L$_{acc}$/L$_{bol}$. However, the larger collection of
data that is available today clearly shows that the majority of
datapoints, which drive the $\dot{M}$-M$_*$ relation, lie well below the k
= 1 limit. To show this we overplot the L$_{acc}$ = L$_{bol}$ line
(i.e. k = 1)  to
the data in Figure~1.

For completeness we have also checked the possibility of a
  straight proportionality between accretion rates and stellar
  mass. A likelihood ratio test allows the rejection of the
  null hypothesis $\dot{M}$ $\propto$M$_*$ to more than six $\sigma$.

\section{$\dot{M}$-$M_*$ as predicted by X-ray photoevaporation}
In the previous section we briefly summarised the available
observations to date and showed that they never yield values of $\alpha$
below 1.55 in the solar mass range. This is difficult to reconcile with the value of
1.35 predicted by Clarke \& Pringle (2006), for a population of discs
dispersed by EUV photoevaporation in the UV switch model (Clarke et
al. 2001, Alexander, Clarke \& Pringle, 2006ab). Following CP06's
argument, the lowest accretion rate measured at a given mass should be
set by the lowest possible photoevaporation rate for the same
mass. Before the onset of photoevaporation the evolution of the
  mass accretion rate follows the usual viscous laws, which predict an
power-law decay with time of $\dot{M}$. Hence young stars spend most of
their time at low accretion rates where one has a higher chance of
observing them. If the lowest allowed accretion rate is determined by
photoevaporation, then this is equivalent to saying that the most
probable observed accretion rate for a given star is $\dot{M}$ =
$\dot{M}$$_{wind}$, where $\dot{M}$$_{wind}$ ist the mass loss rate
due to photoevaporation. For solar mass stars in the UV-switch model this rate is $\sim
10^{10} M_{\odot}/yr$ and scales as the square root of the product of
stellar mass and ionising flux: 

\begin{equation}
\dot{M}_{wind} \propto (M_* \, \phi)^{1/2}
\end{equation}

\noindent giving $\dot{M} \propto M_*^{1.35}$ if the ionising flux simply scales with stellar luminosity. 
If however the UV flux is mainly chromospheric in origin and thus has the same scaling with stellar mass as the 
X-ray luminosity, then
\begin{equation} 
\dot{M} \propto M_* ^{(1+\beta)/2}, 
\end{equation}

\noindent where $\beta$ is the exponent of the X-ray luminosity
function. As will be shown below $\beta$ is roughly 1.7 (Preibisch et
al 2005) for low mass stars, giving again $\dot{M} \propto
M_*^{1.35}$, like in the case where the ionising flux scales with
stellar luminosity .

In recent years several works have shown that X-ray photoevaporation
dominates over EUV photoevaporation for stars with masses of one or
below one solar mass (Ercolano et al. 2009, Owen et al. 2010, 2011, 2012). In the
case of X-ray photoevaporation the mass loss rate, $\dot{M}$$_{wind}$, scales linearly
with the X-ray luminosity, implying that the $\dot{M}$-$M_*$ relation
for a population of discs dispersing via X-ray photoevaporation is
completely determined by the shape of the X-ray luminosity
function. As opposed to Dullemond, Natta \&
Testi, (2006)  and Alexander \& Armitage (2006), this requires no
spread in initial conditions other than the 
dependance on stellar mass. Indeed we argue here that the relation is
primarily driven by the observed accretion rate at late times (just
before dispersal) where the disc spends most of its time. Therefore
the initial conditions are completely irrelevant as they are washed
out after one viscous time. Our model would return the same result
regardless of whether a spread in intial conditions is assumed. 

In Figure 2 we show L$_X$ as function of stellar mass for various
mass ranges in the COUP sample (Preibisch et al. 2005), obtained
with the Chandra X-ray Telescope in the Orion Nebular Cloud (ONC). 
This plot also shows roughly two regimes for the $L_X - M_*$
distribution, where the lower mass stars have a steeper dependence on
stellar mass compared to the higher mass stars. It is indeed well
known that the slope of the distribution flattens out for the higher stellar
masses, where  X-ray production becomes less efficient. The
black solid line shows the power-law slope, $\beta$, for the entire sample ($\beta =
1.28$), the cyan dotted line shows the slope obtained when only stars
with masses lower than 1. M$_{\odot}$ are considered  ($\beta =
1.71$), and the orange dashed line shows the slope for objects in the
higher mass range between 1 and 10 M$_{\odot}$ ($\beta \sim
1.46$). 

It is worth noticing at this point that the slopes quoted include
  the entire sample of accretors and non-accretors in the Preibisch et
  al (2005) data set. According to Preibisch et al (2005), however,
  there are differences in the X-ray luminosities between accretors
  and non-accretors, where non-accretors show marginally higher X-ray
  luminosities that are roughly consistent with those of rapidly
  rotating main sequence stars and they also show a clearer
  correlation with stellar mass, compared to the accretors. The
  accretors, on the other hand, have somewhat suppressed X-ray
  luminosities and the correlation with stellar mass is  
 also not so clear; this is probably due to whatever effect is
 suppressing the X-ray luminosity, which may have to do with the
 presence of  a disc or with whatever is damping the X-ray activity (although
 see Drake et al. 2009 for the opposite interpretation). One has to be
 careful however not to over-interpret this
 discrepancy and it is indeed difficult to estimate any uncertainty on
 the X-ray luminosity function. The main
 problem is that the definition of accretors and non-acccretors used
 by Preibisch et al. (2005)  was based on emission lines and it is
 well known that these can show strong time-variability, leading to
 large uncertainties in the classification. In view of this, and also
 considering the differences between this data-set and the Taurus
 data-set which will be discussed below, we conclude that the
 uncertainties on the quoted value of the slope in the X-ray
 luminosity function are probably larger than those stated here. 

Figure 3 shows the same for the sample of G\"udel et al. (2007)
obtained with the XMM Newton X-ray telescope in the Taurus molecular
cloud. The number of sources in this survey is however much lower and
hence it is more difficult to draw significant statistics,
particularly in the higher mass bin. However qualitatively similar
results are obtained, where $\beta = 1.38, 1.44$ and $1.04$ for the
whole sample, the lower and the higher mass ranges defined above,
respectively. 

The values of $\beta$ obtained for the lower mass objects in the
Preibisch et al (2005) sample compare well
with the values of $\alpha$ obtained for the same mass range, which is
what one would expect if disc dispersal in this stellar mass range is
dominated by X-ray photoevaporation. The same process is not expected
to be the dominant disc dispersal mechanism for discs around high mass
stars, where X-ray production is expected to be lower and the
corresponding photoevaporation rates then too weak to compete with the
higher accretion rates. In this context the lack of agreement between
$\alpha$ and $\beta$ in the higher mass range is not surprising. 

It is difficult to speculate what the dominant dispersal
  mechanism at higher masses may be. The drop in the X-ray luminosities
  for these higher mass stars, implies that if photoevaporation is
  still the main driver of dispersal, the main heating source must be
  EUV or FUV photons. Detailed hydrodynamical wind solutions for these
  objects have yet to be calculated, although some estimates using a
  simpler approach were provided by Gorti, Dullemond \& Hollenbach
  (2009), which show that photoevaporation by FUV radiation may be a
  viable solution for the fast dispersal of discs around higher mass stars.

If X-ray photoevaporation is indeed controlling disc dispersal around
low mass stars, hence determining the slope of the $\dot{M}$-M$_*$
relation, one other issue to be considered is the lowest possible accretion rate mesurable at a
given mass in a sample of discs that are dispersed by X-ray
photoevaporation. Owen et al (2010) show that the final phase of rapid
disc dispersal begins roughly when the accretion rates become about a factor
ten lower than the photoevaporation rates. For the ONC sources shown
in Figure 1, the lowest X-ray luminosities for solar mass
stars are of the order of approximately 10$^{29}$ ergs/s. This
corresponds roughly to X-ray photoevaporation rates of
10$^{-9}$M$_{\odot}$/yr, implying that the lowest accretion rates that
should be measurable are of order  10$^{-10}$M$_{\odot}$/yr, which is
consistent with observational data in this mass bin (Manara et al 2012). 

A final, perhaps more stringent test of this model is a
  comparison of the $L_X$-$M_*$ distribution against the $\dot{M}$-$M_*$
  distribution. The one-to-one mapping of the wind mass loss rate with
the X-ray luminosity would indeed suggest that their normalised
distribution in the low mass bin, where X-ray photoevaporation
dominates, should be indistinguishable. Unfortunately a direct comparison of the data-sets is impossible 
since the distributions of stellar masses in the two samples (even in
bins around solar-type stars) is formally different to very high
significance. The likely cause of this is that our $\dot{M}$-$M_*$ sample
contains a large number of objects from Taurus, which is known to have
an unusual IMF (Luhman 2004). Therefore, in order to perform a meaningful statistical test we need to resample both distributions onto the same underlying mass-function.

\begin{figure}
\centering
\includegraphics[width=0.5\textwidth]{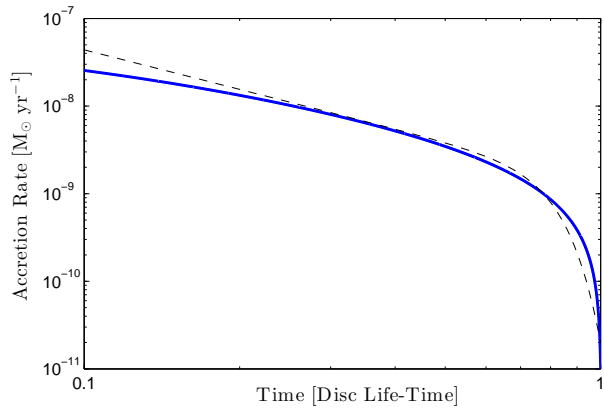}
\caption{Comparison between a full viscous calculation (solid) and the simple formula (dashed)}\label{fig:mdot_compare}
\end{figure}

Thus we construct a statistical test to see if we can rule out the
null-hypothesis that the observed $\dot{M}$ distribution is purely driven
by disc accretion terminated by X-ray photoevaporation. We choose the
mass range (0.2-1.2M$\odot$) where X-ray photoevaporation is likely to
be dominant (e.g. Owen et al. 2012). In this mass-range we then
randomly sample both our $\dot{M}$-$M_*$ and $\dot{M}$-L$_X$ distributions onto a
Chabrier IMF (Chabrier 2003) where the new resampled distributions
consist of 100 data points (roughly the maximum number possible before
random noise is the dominant source of variation). We then convert our
$\dot{M}$-L$_X$ distribution into a $\dot{M}$-M$_*$ distribution by assuming that
the accretion rate follows a simple viscous disc model of a $t^{-3/2}$
decline in accretion rate until the accretion rate equals the
mass-loss rate, where we use the Owen et al. (2012) fitting function:
\begin{equation}
\dot{M}_w=6.25\times10^{-9}\left(\frac{L_X}{10^{30} {\rm \, erg s}^{-1}}\right)^{1.14}
\end{equation}
but ignore the very weak ($\sim M_*^{-0.068}$) stellar mass dependance. At this point the accretion rate follows an exponential cut-off with a time-scale approximately 10\% of the disc's lifetime. Formally the expression we use is given by:
\begin{equation}
\dot{M}_*\propto t^{-3/2}\exp\left[-\left(\frac{t}{\tau_{\rm disc}}\right)^7\right]\label{eqn:mdot}
\end{equation} 
where $\tau_{\rm disc}$ is a scale time which modifies the disc's
    life-time so the exponential cut-off begins when the viscous
    disc's accretion rate drops below the wind rate.  Such an
    evolution does not contain information about the wind profile, as
    it is just matching together two different phases of disc
    evolution (primordial disc evolution \& inner disc draining, see Owen
    et al. 2010 for a discussion)\footnote{Note the form of the accretion
    rate evolution looks very similar to
    the semi-analytic solutions presented by Ruden (2004) using a
    Green's function approach for the EUV wind}. Formally this
  exercise is completed in a scale-free manner (allowing us to ignore
  unconstrained disc parameters e.g. $\alpha$), with no explicit
  choice of what $\tau_{\rm disc}$ is. In order to find the lifetime
  $\tau_{\rm disc}$, we find the time $\tau_{\rm disc}$ so that $\dot{M} (t) =
  \dot{M}_w (L_X)$. For simplicity, we approximate the dependence of
  $\dot{M}$ with $t^{-3/2}$, so that the equation can be solved
  analytically. 
 We note that this
  construction is consistent with providing the observed spread in
  disc lifetimes from the spread in X-ray luminosity alone, as
  demonstrated in Owen et al. (2011). A comparison between this
formula and an actual viscous calculation from Owen et al. (2011) is
shown in Figure~\ref{fig:mdot_compare}. 

Such a formula provides an approximate description of the disc's
accretion rate evolution. The $\dot{M}$-L$_X$ distribution is thus convolved with this
expression in order to produce an $\dot{M}$-$M_*$ relation which can then
be compared with the observed $\dot{M}$-$M_*$ relation. Thus with our two
re-sampled $\dot{M}$-M$_*$ relations (one observed, one calculated from the
M$_*$-L$_X$) we calculated the Kaplan-Meier distributions and perform a
null-hypothesis test to determine whether we can reject the
null-hypothesis that these two distributions are different. Since
we are re-sampling our two distributions onto the same underlying mass
distribution we perform 500 realisations of this random samplings to
get a sense of the re-sampling error. In Figure~5, we
show the Kaplan-Meier distributions resulting from 10 re-sampling
realisations for the observed M$_*$-$\dot{M}$ (solid) and the M$_*$-$\dot{M}$
distributions calculated from the observed M$_*$-L$_X$ distribution
(dashed). Furthermore, in Figure~\ref{fig:p_values} we show a
histogram of the P-values resulting from our hypothesis tests (both
using the log-rank and Gehan methods - see Feigelson \& Nelson 1985)
for each of our random realisations,
where the P value is an estimate of the probability that the two
distributions are drawn from the same underlying sample. 
Therefore, they are unlikely to come from the same
underlying distribution if P is small ($<$0.05 $\sim$2$\sigma$). Given that in
our case P is clearly generally not small one cannot rule out the
null hypothesis. This is striking given that the two distributions should
be independent unless they are connected by the
photoevaporation model. 
In summary, both figures show extremely good agreement between the two
distributions and we are unable to reject the null hypothesis,
suggesting that the two populations are indeed connected by the X-ray
photoevaporation model.

\begin{figure}
\centering
\includegraphics[width=\columnwidth]{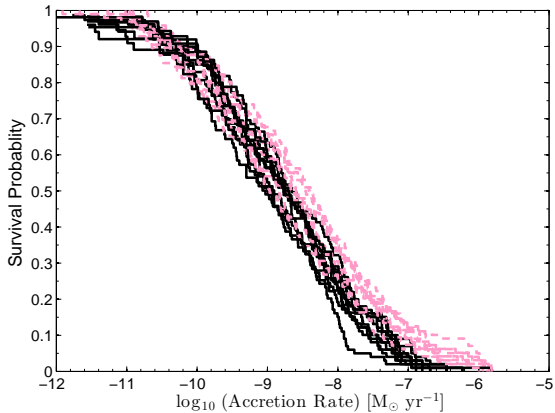}
\caption{KM distributions for the re-sampled $\dot{M}$-M$_*$ (solid) and calculate $\dot{M}$-M$_*$ (dashed) relations.}\label{KM_dist}
\end{figure}

\begin{figure}
\centering
\includegraphics[width=\columnwidth]{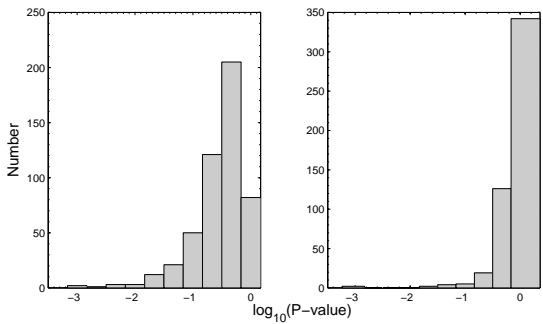}
\caption{P-values from the hypothesis test: log-rank (left), Gehan (right)}\label{fig:p_values}
\end{figure}

We note, further, that the disagreement is very small, and mostly evident at high accretion
rates which are likely dominated by accretion rates calculated from
flaring X-ray values. Furthermore the simple viscous model is likely
to break down at early times due to variability. 

The X-ray luminosity data then shows, in summary, that the observed
$\dot{M}$-M$_*$ relation in pre-main-sequence stars is consistent with being
a simple consequence of disc dispersal by X-ray photoevaporation. 

\section{Summary}

 We have presented a statistical analysis of accretion rates of
pre-main sequence stars as a function of stellar mass in order to
establish whether the steep relation between accretion rates and mass
may be a consequence of selection or detection thresholds, as feared
in past (e.g. Clarke \& Pringle, 2006; Tilling et al 2008). With the
large amount of data available we show (using survival statistics)
that selections/detection biases are {\it not} driving the $\dot{M}-M_*$ 
relation.

We find the slope of the power law relation to be between $\sim$1.6
and 1.9 for stars with masses lower that 1 $M_{\odot}$. Such slopes
are similar to the slopes observed in the X-ray luminosity function of
young stars in near-by clusters (e.g. Preibisch et al, 2005; G\"udel
et al 2007). We show that X-ray photoevaporation predicts indeed that
the observed $\dot{M}-M_*$ relation should be completely determined by
the X-ray luminosity of the stars, which thus imprints a signature on
the observed accretion rates distribution in a given
cluster. 

Furthermore we demonstrate that a synthetic $\dot{M}-M_*$
dataset constructed from the X-ray luminosity function of young stars
in the Orion Nebular Cluster (Preibisch et al. 2005) is statistically
indistinguishable from the observed $\dot{M}-M_*$ dataset, hence
lending further support that discs around young stars disperse
predominantly by X-ray photoevaporation. 

\section{Acknowledgements}

We thank the anonymous referee for a positive and constructive
report. The calculations were partially performed on the Sunnyvale cluster at CITA which is funded by the Canada Foundation for Innovation.

\onecolumn
\begin{longtable}{|lrrll|}
\caption{Regions and references of aquired sample} \\
\hline
Region & \multicolumn{1}{l}{\# measurements} & \multicolumn{1}{l}{\# upperlimits} & methods\footnotemark & references \\ \hline
\endfirsthead

\caption{continued} \\
\hline 
Region & \multicolumn{1}{l}{\# measurements} & \multicolumn{1}{l}{\# upperlimits} & methods & references \\ \hline
\endhead

\hline \multicolumn{5}{|c|}{{Continued on next page}} \\ \hline
\endfoot

\hline
\endlastfoot
ASSOC II SCO + ASS Sco OB + & 10 & 4 & a,b,c & total \\
LCC + USco & 1 & 0 & a,b & Curran et al. 2011 \\*						
& 4 & 3 & a,c & Herczeg et al. 2009 \\*						
& 0 & 1 & a & Mohanty et al. 2005 \\*						
& 5 & 1 & a & Garcia Lopez et al. 2006 \\ \hline				
Cra Dark Cloud & 2 & 3 & a & total \\*  
& 2 & 2 & a &  Fang et al. 2013\\*						
& 0 & 1 & a &  Garcia Lopez et al. 2006\\ \hline				
Cep OB2 Tr37 + Cep OB2 NGC7160 & 90 & 10 & d & total \\*  
& 81 & 1 & d &  Sicilia-Aguilar et al. 2010\\* 					
& 35 & 27 & d &  Sicilia-Aguilar et al. 2006\\ \hline 				
Cha I + Cha II & 82 & 15 & a,d,e & total\\* 
& 2 & 7 & a,e &  Natta et al. 2004\\* 						
& 6 & 12 & e &  Muzerolle et al. 2005\\* 					
& 10 & 1 & a &  Costigan et al. 2012\\*						
& 4 & 0 & a &  Fang et al. 2013\\* 						
& 16 & 0 & d &  Hartmann et al. 1998\\* 					
& 0 & 2 & a &  Mohanty et al. 2005\\* 						
& 18 & 0 & a &  Robberto et al. 2012\\* 					
& 37 & 0 & a &  Biazzo et al. 2012\\ \hline					
IC 348 & 19 & 3 & a,e,f & total \\*  						
& 6 & 0 & e &  Muzerolle et al. 2003\\* 					
& 4 & 2 & a &  Mohanty et al. 2005\\*						
& 13 & 1 & a,f &  Dahm 2008\\ \hline 						
L1630N & 70 & 0 & a & Fang et al. 2009 \\ \hline 				
L1641 & 103 & 0 & a & Fang et al. 2009 \\ \hline				
Lupus & 5 & 2 & a,b,c & total \\* 				
& 3 & 2 & a &  Fang et al. 2013\\* 						
& 1 & 0 & a,b &  Curran et al. 2011\\* 						
& 1 & 0 & c &  Herczeg \& Hillenbrand 2008\\* 					
& 1 & 0 & a &  Garcia Lopez et al. 2006\\ \hline				
Orion Nebular Cluster & 698 & 85 & a,g & Manara et al. 2012 \\ \hline	
TWA & 6 & 2 & a,b,c,e & total \\* 
& 1 & 1 & c,e &  Muzerolle et al. 2000\\* 					
& 1 & 0 & a &  Donati et al. 2011a\\*						
& 2 & 0 & a,b &  Curran et al. 2011\\* 						
& 3 & 0 & c,e &  Herczeg \& Hillenbrand 2008\\* 				
& 4 & 1 & c &  Herczeg et al. 2009\\* 						
& 0 & 1 & a &  Mohanty et al. 2005\\ \hline					
$\sigma$ Orionis & 39 & 45 & a,c,d & total \\* 					
& 6 & 2 & c &  Rigliaco et al. 2012\\* 						
& 12 & 23 & a & Gatti et al. 2008\\* 						
& 30 & 30 & d & Rigliaco et al. 2011\\ \hline					
$\epsilon$ Cha & 4 & 0 & a & Fang et al. 2013 \\ \hline				
Taurus & 118 & 40 & a,b,c,d,e,f & total \\* 						
& 17 & 0 & c &  Gullbring et al. 1998\\*					
& 6 & 1 & c,e &  Muzerolle et al. 2003\\* 					
& 9 & 6 & e &  Muzerolle et al. 2005\\* 					
& 55 & 25 & f &  White et al. 2001\\* 						
& 40 & 0 & c,d &  Hartmann et al. 1998\\* 					
& 2 & 0 & a,b &  Curran et al. 2011\\* 						
& 16 & 2 & c,e &  Herczeg \& Hillenbrand 2008\\* 				
& 6 & 3 & a &  Mohanty et al. 2005\\* 						
& 1 & 0 & f &  White et al. 2005\\* 						
& 3 & 7 & f &  White et al. 2003\\* 						
& 23 & 3 & f &  White et al. 2004\\ \hline 					
$\rho$ Oph & 46 & 72 & a,e & total \\* 						
& 46 & 71 & a &  Natta et al. 2006\\* 						
& 7 & 3 & a,e &  Natta et al .2004\\* 						
& 13 & 3 & a &  Gatti et al. 2006\\* 						
& 1 & 0 & a &  Donati et al. 2011b\\*						
& 1 & 0 & a &  Curran et al. 2011\\* 						
& 1 & 1 & a & Mohanty et al. 2005\\ \hline					
various & 28 & 7 & a,b & total \\*
& 6 & 3 & a & Fang et al. 2013\\*						
& 1 & 0 & a & Mohanty et al. 2005 \\*						
& 20 & 4 & a & Garcia Lopez et al. 2006	\\*					
& 1 & 0 & a &  Donati et al. 2011c\\*						
& 1 & 0 & a,b &  Curran et al. 2011  						
\footnotetext[1]{a: Emission lines luminosity converted to L$_{acc}$ using empirical calibrations  \newline b: X-ray emission (Curran et al. 2011) \newline c: blue continuum excess measured spectroscopically \newline d: U-band excess measured photometrically  \newline  e: fit of H$\alpha$-profile  \newline f: veiling measurements of photospheric lines \newline g: U-band
excess using the two-colors diagram (Manara et al. 2012)} 
\end{longtable}

\twocolumn

\label{lastpage}


\begin{thebibliography}{99}


\bibitem[Alexander 
\& Armitage(2006)]{2006ApJ...639L..83A} Alexander, R.~D., \& Armitage, P.~J.\ 2006, \apjl, 639, L83 

\bibitem[Alexander et al.(2006)]{2006MNRAS.369..216A} Alexander, R.~D., 
Clarke, C.~J., \& Pringle, J.~E.\ 2006, \mnras, 369, 216 


\bibitem[Alexander et al.(2006)]{2006MNRAS.369..229A} Alexander, R.~D., 
Clarke, C.~J., \& Pringle, J.~E.\ 2006, \mnras, 369, 229 

\bibitem[Antoniucci et 
al.(2011)]{2011A&A...534A..32A} Antoniucci, S., Garc{\'{\i}}a L{\'o}pez, R., Nisini, B., et al.\ 2011, \aap, 534, A32 

\bibitem[Baraffe et 
al.(1998)]{1998A&A...337..403B} Baraffe, I., Chabrier, G., Allard, F., \& Hauschildt, P.~H.\ 1998, \aap, 337, 403 


\bibitem[Biazzo et 
al.(2012)]{2012A&A...547A.104B} Biazzo, K., Alcal{\'a}, J.~M., Covino, E., et al.\ 2012, \aap, 547, A104 


\bibitem[Calvet et al.(2004)]{2004AJ....128.1294C} Calvet, N., Muzerolle, 
J., Brice{\~n}o, C., et al.\ 2004, \aj, 128, 1294 

\bibitem[Chabrier(2003)]{2003ApJ...586L.133C} Chabrier, G.\ 2003, \apjl, 
586, L133 

\bibitem[Clarke et al.(2001)]{2001MNRAS.328..485C} Clarke, C.~J., Gendrin, 
A., \& Sotomayor, M.\ 2001, \mnras, 328, 485 

\bibitem[Clarke 
\& Pringle(2006)]{2006MNRAS.370L..10C} Clarke, C.~J., \& Pringle, J.~E.\ 2006, \mnras, 370, L10 


\bibitem[Costigan et al.(2012)]{2012MNRAS.427.1344C} Costigan, G., Scholz, 
A., Stelzer, B., et al.\ 2012, \mnras, 427, 1344 


\bibitem[Curran et 
al.(2011)]{2011A&A...526A.104C} Curran, R.~L., Argiroffi, C., Sacco, G.~G., et al.\ 2011, \aap, 526, A104 

\bibitem[D'Antona 
\& Mazzitelli(1997)]{1997MmSAI..68..807D} D'Antona, F., \& Mazzitelli, I.\ 1997, MEMSAI, 68, 807 

\bibitem[Dahm(2008)]{2008AJ....136..521D} Dahm, S.~E.\ 2008, \aj, 136, 521 

\bibitem[Donati et al.(2011b)]{2011MNRAS.412.2454D} Donati, J.-F. et al.\ 2011b, \mnras, 412, 2454

\bibitem[Donati et al.(2011c)]{2011MNRAS.417.1747D} Donati, J.-F., Gregory, 
S.~G., Montmerle, T., et al.\ 2011c, \mnras, 417, 1747 

\bibitem[Donati et al.(2011a)]{2011MNRAS.417..472D} Donati, J.-F., Gregory, 
S.~G., Alencar, S.~H.~P., et al.\ 2011a, \mnras, 417, 472 

\bibitem[Dullemond et al.(2006)]{2006ApJ...645L..69D} Dullemond, C.~P., 
Natta, A., \& Testi, L.\ 2006, \apjl, 645, L69 

\bibitem[Ercolano et al.(2009)]{2009ApJ...699.1639E} Ercolano, B., Clarke, 
C.~J., \& Drake, J.~J.\ 2009, \apj, 699, 1639 


\bibitem[Fang et 
al.(2009)]{2009A&A...504..461F} Fang, M., van Boekel, R., Wang, W., et al.\ 2009, \aap, 504, 461 

\bibitem[Fang et 
al.(2013)]{2013A&A...549A..15F} Fang, M., van Boekel, R., Bouwman, J., et al.\ 2013, \aap, 549, A15 


\bibitem[Feigelson 
\& Nelson(1985)]{1985ApJ...293..192F} Feigelson, E.~D., \& Nelson, P.~I.\ 1985, \apj, 293, 192 

\bibitem[Garcia Lopez et 
al.(2006)]{2006A&A...459..837G} Garcia Lopez, R., Natta, A., Testi, L., \& Habart, E.\ 2006, \aap, 459, 837 


\bibitem[Gatti et 
al.(2006)]{2006A&A...460..547G} Gatti, T., Testi, L., Natta, A., Randich, S., \& Muzerolle, J.\ 2006, \aap, 460, 547 

\bibitem[Gullbring et al.(1998)]{1998ApJ...492..323G} Gullbring, E., 
Hartmann, L., Briceno, C., \& Calvet, N.\ 1998, \apj, 492, 323 

\bibitem[Gatti et 
al.(2008)]{2008A&A...481..423G} Gatti, T., Natta, A., Randich, S., Testi, L., \& Sacco, G.\ 2008, \aap, 481, 423 


\bibitem[Gregory et al.(2006)]{2006MNRAS.373..827G} Gregory, S.~G., 
Jardine, M., Collier Cameron, A., \& Donati, J.-F.\ 2006, \mnras, 373, 827 


\bibitem[G{\"u}del et 
al.(2007)]{2007A&A...468..353G} G{\"u}del, M., Briggs, K.~R., Arzner, K., et al.\ 2007, \aap, 468, 353 

\bibitem[Hartmann et al.(1998)]{1998ApJ...495..385H} Hartmann, L., Calvet, 
N., Gullbring, E., \& D'Alessio, P.\ 1998, \apj, 495, 385 

\bibitem[Herczeg et al.(2009)]{2009ApJ...696.1589H} Herczeg, G.~J., Cruz, 
K.~L., \& Hillenbrand, L.~A.\ 2009, \apj, 696, 1589 

\bibitem[Herczeg 
\& Hillenbrand(2008)]{2008ApJ...681..594H} Herczeg, G.~J., \& Hillenbrand, L.~A.\ 2008, \apj, 681, 594 


\bibitem[Luhman(2004)]{2004ApJ...617.1216L} Luhman, K.~L.\ 2004, \apj, 617, 
1216 


\bibitem[Manara et al.(2012)]{2012ApJ...755..154M} Manara, C.~F., Robberto, 
M., Da Rio, N., et al.\ 2012, \apj, 755, 154 


\bibitem[Manara et 
al.(2013)]{2013A&A...551A.107M} Manara, C.~F., Testi, L., Rigliaco, E., et al.\ 2013, \aap, 551, A107 

\bibitem[Mohanty et al.(2005)]{2005ApJ...626..498M} Mohanty, S., 
Jayawardhana, R., \& Basri, G.\ 2005, \apj, 626, 498 


\bibitem[Muzerolle et al.(2000)]{2000ApJ...545L.141M} Muzerolle, J., 
Brice{\~n}o, C., Calvet, N., et al.\ 2000, \apjl, 545, L141 


\bibitem[Muzerolle et al.(2003)]{2003ApJ...592..266M} Muzerolle, J., 
Hillenbrand, L., Calvet, N., Brice{\~n}o, C., 
\& Hartmann, L.\ 2003, \apj, 592, 266 

\bibitem[Muzerolle et al.(2005)]{2005ApJ...625..906M} Muzerolle, J., 
Luhman, K.~L., Brice{\~n}o, C., Hartmann, L., 
\& Calvet, N.\ 2005, \apj, 625, 906 


\bibitem[Natta et 
al.(2004)]{2004A&A...424..603N} Natta, A., Testi, L., Muzerolle, J., et al.\ 2004, \aap, 424, 603 

\bibitem[Natta et 
al.(2006)]{2006A&A...452..245N} Natta, A., Testi, L., \& Randich, S.\ 2006, \aap, 452, 245 


\bibitem[Owen et al.(2010)]{2010MNRAS.401.1415O} Owen, J.~E., Ercolano, B., 
Clarke, C.~J., \& Alexander, R.~D.\ 2010, \mnras, 401, 1415 


\bibitem[Owen et al.(2011)]{2011MNRAS.412...13O} Owen, J.~E., Ercolano, B., 
\& Clarke, C.~J.\ 2011, \mnras, 412, 13 

\bibitem[Owen et al.(2012)]{2012MNRAS.422.1880O} Owen, J.~E., Clarke, 
C.~J., \& Ercolano, B.\ 2012, \mnras, 422, 1880 


\bibitem[Padoan et al.(2005)]{2005ApJ...622L..61P} Padoan, P., Kritsuk, A., 
Norman, M.~L., \& Nordlund, {\AA}.\ 2005, \apjl, 622, L61 

\bibitem[Palla 
\& Stahler(1999)]{1999ApJ...525..772P} Palla, F., \& Stahler, S.~W.\ 1999, \apj, 525, 772 

\bibitem[Preibisch et al.(2005)]{2005ApJS..160..401P} Preibisch, T., Kim,
Y.-C., Favata, F., et al.\ 2005, \apjs, 160, 401


\bibitem[Rigliaco et 
al.(2011)]{2011A&A...525A..47R} Rigliaco, E., Natta, A., Randich, S., Testi, L., \& Biazzo, K.\ 2011, \aap, 525, A47 


\bibitem[Robberto et al.(2012)]{2012AJ....144...83R} Robberto, M., Spina, 
L., Da Rio, N., et al.\ 2012, \aj, 144, 83 


\bibitem[Sicilia-Aguilar et al.(2010)]{2010ApJ...710..597S} 
Sicilia-Aguilar, A., Henning, T., \& Hartmann, L.~W.\ 2010, \apj, 710, 597 

\bibitem[Siess et 
al.(2000)]{2000A&A...358..593S} Siess, L., Dufour, E., \& Forestini, M.\ 2000, \aap, 358, 593 

\bibitem[Throop 
\& Bally(2008)]{2008AJ....135.2380T} Throop, H.~B., \& Bally, J.\ 2008, \aj, 135, 2380 


\bibitem[Tilling et al.(2008)]{2008MNRAS.385.1530T} Tilling, I., Clarke, 
C.~J., Pringle, J.~E., \& Tout, C.~A.\ 2008, \mnras, 385, 1530 


\bibitem[White 
\& Ghez(2001)]{2001ApJ...556..265W} White, R.~J., \& Ghez, A.~M.\ 2001, \apj, 556, 265 

\bibitem[White 
\& Hillenbrand(2005)]{2005ApJ...621L..65W} White, R.~J., \& Hillenbrand, L.~A.\ 2005, \apjl, 621, L65 

\bibitem[White 
\& Basri(2003)]{2003ApJ...582.1109W} White, R.~J., \& Basri, G.\ 2003, \apj, 582, 1109 

\bibitem[White 
\& Hillenbrand(2004)]{2004ApJ...616..998W} White, R.~J., \& Hillenbrand, L.~A.\ 2004, \apj, 616, 998 


\end{thebibliography}
\end{document}